# Zero-cell photonic crystal nanocavity laser with quantum dot gain

Masahiro Nomura[1, a)], Yasutomo Ota[1, 2], Naoto Kumagai[1], Satoshi Iwamoto[1, 2], and Yasuhiko Arakawa[1, 2]

[1]*Institute for Nano Quantum Information Electronics, The University of Tokyo, 4-6-1 Komaba, Meguro-ku, Tokyo 153-8505, Japan*
[2]*Institute of Industrial Science, The University of Tokyo, 4-6-1 Komaba, Meguro-ku, Tokyo 153-8505, Japan*

We demonstrate laser oscillation in a hexagonal-lattice photonic crystal nanocavity using an InGaAs quantum dot gain material by optical pumping at 5 K. The cavity comprises a defect created by shifting several air holes in a two-dimensional photonic crystal slab structure without removing any air holes to achieve both small mode volume and a high cavity quality factor. The measured cavity quality factors and estimated mode volume for the nanocavity are ~33,000 and ~0.004 $\mu m^3$ ($0.23(\lambda_0/n)^3$). The laser threshold is compared between the zero-cell and L3-type nanocavity lasers, and the zero-cell nanolasers are found to have small thresholds of about one-third of the L3-type nanolasers. This result suggests that a higher Purcell factor of the zero-cell nanolaser is reflected as a smaller laser threshold.

Strong photon confinement drastically enhances the intensity of the electric field per single photon and enables strong light–matter interaction. Since the report on the artificial control of the spontaneous emission probability by E. M. Purcell in 1946,[1)] cavity quantum electrodynamics (C-QED) has been a major source of inspiration for researches in which the interaction between light and materials is manipulated. Since the 1990s, a semiconductor microcavity structure has been a good platform for C-QED research.[2)] The key structural parameters of C-QED systems are the cavity quality factor ($Q$) and the modal volume ($V_m$). High-$Q$ semiconductor microcavities have realized various characteristic physics such as efficient laser oscillation in the weak-coupling regime,[3-10)] vacuum Rabi splitting in the strong-coupling regime,[11-14)] and related physics[15, 16)] depending on the interaction strength. Here, we focus on the weak-coupling regime and realization of lasing with a low laser threshold using a high-$Q/V_m$ cavity.

A zero-cell photonic crystal (PhC) nanocavity has much smaller $V_m$ than the defect-type cavities produced via the removal of air holes and thus a higher Purcell factor ($F_p$) is realizable. So far, hexagonal, referred to as H0-type,[17)] and square–lattice[18)] zero-cell PhC cavities have been designed and lasing has been demonstrated with quantum–well gain materials.[18, 19)] However, if the gain material can be replaced by quantum dots (QDs), then the laser oscillation starts at a much lower pumping power. In addition, a low-density QD system enables a straightforward discussion on the impact of the C-QED effect on the

a)Electronic mail: nomura@iis.u-tokyo.ac.jp

threshold of high-$F_p$ nanolasers at cryogenic temperatures. Therefore, a QD-based zero-cell nanolaser is an attractive system for exploring fundamental physics and realizing practical nanolasers.

In this letter, we demonstrate the first laser oscillation in a zero-cell PhC nanocavity using InGaAs/GaAs QDs for the optical gain under continuous-wave optical pumping at 5 K. Lasing was observed in cavities with cavity $Q$ of ~30,000–40,000 and $V_m$ of ~0.004 μm$^3$ by under the non-resonant coupling condition using self-tuned QD gain[5]. The laser thresholds are compared between the zero-cell and L3-type[20] PhC nanocavity lasers under similar conditions.

Samples were grown by molecular beam epitaxy on an undoped (100)-oriented GaAs substrate. First, a 300-nm-thick GaAs buffer layer was deposited on the substrate and then, a 1-μm-thick $Al_{0.7}Ga_{0.3}As$ sacrificial layer was grown. Finally, a 130-nm-thick GaAs slab layer along with a single self-assembled InGaAs QD layer was grown at the center of the slab. The photoluminescence (PL) peak of the QD ensemble was observed at 980 nm at 5 K. The nominal areal QD density was ~4 × 10$^8$ cm$^{-2}$. We investigated an air-clad two-dimensional PhC nanocavity with a hexagonal air hole lattice. The studied H0-type nanocavity comprises a defect created by shifting several air holes in a two-dimensional PhC slab structure without removing any air holes as shown in Fig. 1(a). The radius of an air hole is ~ 0.25$a$ ($a$: lattice constant of the PhC). The details of the fabrication process can be found in our previous report.[10] The design of the cavity is based on our previous study employing three-dimensional finite-difference time domain (FDTD) simulations.[21] The shifts of on-axis air holes are $S_{1x} = 0.14a$, $S_{2x} = 0$, $S_{3x} = 0.06a$, $S_{1y} = 0.04a$, and $S_{2y} = 0.02a$. The calculated cavity $Q$ and $V_m$ of the fundamental mode are ~2 × 10$^5$ and ~0.004 μm$^3$ (0.23($\lambda_0/n$)$^3$), respectively. The calculated $V_m$ is about one-third of that of an L3-type nanocavity, and thus a larger $F_p$ is expected. Fig. 1(b) shows the spatial distributions of $E_y$, which is the main electric field component of the investigated cavity mode.

The measurements were performed at 5 K using a micro-PL setup. A continuous-wave Ti:sapphire laser was used as an excitation source. The excitation wavelength was set to 880 nm to generate photocarriers in the wetting layer (PL peak of 880 nm) to suppress unnecessary background-emitter pumping. An excitation beam was focused on the surface of the sample using a microscope objective lens (40×, numerical aperture of 0.6) in the normal direction. The theoretical diameter of the excitation spot formed on the surface of the sample was calculated to be ~2 μm.

Figure 2(a) shows the PL spectrum of the QDs in the unpatterned region of the sample with 700 nW pumping. The QDs provide gain to the cavity mode and realize laser oscillation through the non-resonant coupling process.[22] Figure 2(b) shows PL spectra measured for H0-type cavities with different $a$ values between 290 and 305 nm. Laser oscillation was observed in a broad wavelength range, where the QDs could provide sufficient optical gain with cavities having $Q$ values of ~30,000–40,000.

We chose a typical H0-type cavity with $a$ = 300 nm to investigate laser characteristics. Figure 2(c) shows the PL spectra of the fundamental mode of the H0-type cavity at excitation powers of 80 nW (left panel) and 10 μW (right panel). The blue balls represent the experimental results and the red lines represent curves fitted using Voigt functions and taking account of the 16-pm-spectral resolution of our detection system. Figures 22(d) shows the light-in versus light-out (L–L) plot. As the excitation power increases, the cavity mode has a super-linear increase in the output power with a gentle s-shaped L–L plot. This type of smooth transition from the thermal to the stimulated emission region is typically observed in

high-$\beta$ lasers in which spontaneous emission efficiently couples with the lasing mode.[23] The laser threshold, which can be defined as the point of inflection, is ~300 nW. We observed the gradual blueshift of the peak caused by the carrier-induced refractive index change with negative value.[24] The value of cavity $Q$ is estimated from the linewidth and found to be 33,000 ± 1,000 at the transparent condition. Linewidth narrowing above the threshold is also observed as the pump power increases. The observed linewidth narrowing is ~5% at $10P_{th}$. This amount seems reasonable compared with a micropillar laser using a resonantly coupled single QD as the optical gain (~10% narrowing at $10P_{th}$).[9] We also constructed an L–L plot for excitation at a wavelength of 800 nm. In this case, photocarriers are generated in the whole GaAs slab. The laser threshold is ~10 nW, which is 30 times less than that for the wetting–layer excitation. The thicknesses for the interaction of the excitation light and absorptive materials are ~0.45 nm (thickness of the wetting layer) and 130 nm (slab thickness) for the 880-nm- and 800-nm-pumping cases. The number of photo-carriers differs by a factor of ~300, but the threshold differs only by a factor of 30. This comparison indicates that the wetting layer excitation is ~10 times more efficient than the GaAs layer excitation. This is mainly because the photocarriers are generated near the QDs and efficiently trapped by them. This efficient use of photocarriers results in less undesirable background emission, less heating, and a smaller carrier-induced peak shift etc.

An H0-type cavity has smaller mode volume than other PhC cavities investigated thus far. We conducted comparative measurements of a laser threshold between the H0-type ($V_m \sim 0.23(\lambda_0/n)^3$) and the L3-type ($V_m \sim 0.7(\lambda_0/n)^3$) PhC nanocavities. A smaller value of $V_m$ for an H0-type nanocavity results in a higher vacuum field amplitude. The enhanced vacuum field amplitude increases the cavity photon emission rate and leads to a lower laser threshold. We fabricated H0-type and L3-type PhC nanocavities on the same wafer and measured the laser thresholds of several cavities, which were observed to have a cavity peak at 950 ± 5 nm and $Q$ = 21,000 ± 3,000. The plots of the laser threshold values obtained are shown in Fig. 3. There are some variations in the threshold, however, H0-type nanocavities had lower threshold values than L3-type nanocavities. The average threshold values obtained are 280 and 780 nW for H0-type and L3-type cavities, respectively. The cavities are not specifically coupled to the QDs, but using non-resonant coupling gain and/or background continuum gain. The variations in the threshold could be due to the difference in the number of non-resonantly coupled QDs and the difference in the carrier trapping rate of the QDs. The lower averaged threshold for the H0-type nanolasers, about one third of that of the L3-type nanolasers, can be attributed to strong Purcell enhancement of the cavity photon emission rate.

In summary, QD-based zero-cell PhC nanocavity lasers have been demonstrated for the first time. Laser oscillation was observed in nanocavities with a cavity $Q$ ~33,000 and $V_m$ ~0.004 μm³ $(0.23(\lambda_0/n)^3)$ with continuous-wave optical pumping at 5 K. A comparative measurement of the laser threshold between two nanolasers with different mode volumes indicates that the obtained lower threshold of the zero-cell nanocavity results from a higher Purcell factor induced by the smaller mode volume.

We thank S. Ishida, S. Ohkouchi, M. Shirane, Y. Igarashi, K. Watanabe and, R. Ota for their technical support and fruitful discussions. This research was supported by the Special Coordination Funds for Promoting Science and Technology, Ministry of Education, Culture, Sports, Science and Technology, Japan.

Figures and figure captions

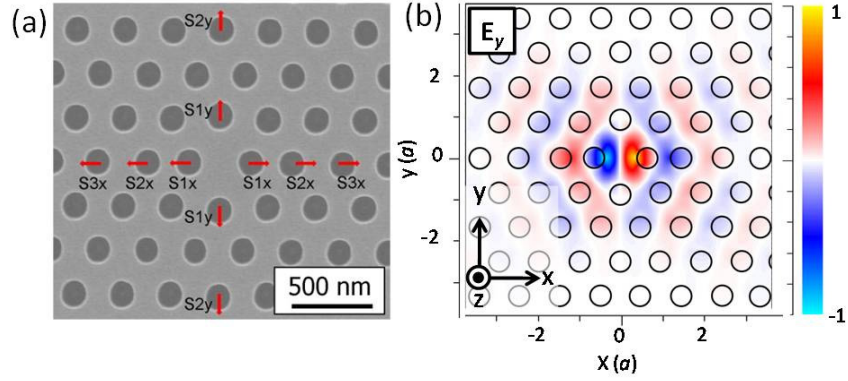

FIG. 1. (a) SEM image of a hexagonal-lattice zero-cell-type (H0-type) PhC nanocavity. (b) Three-dimensional FDTD simulated profile of the $E_y$ component for the fundamental mode.

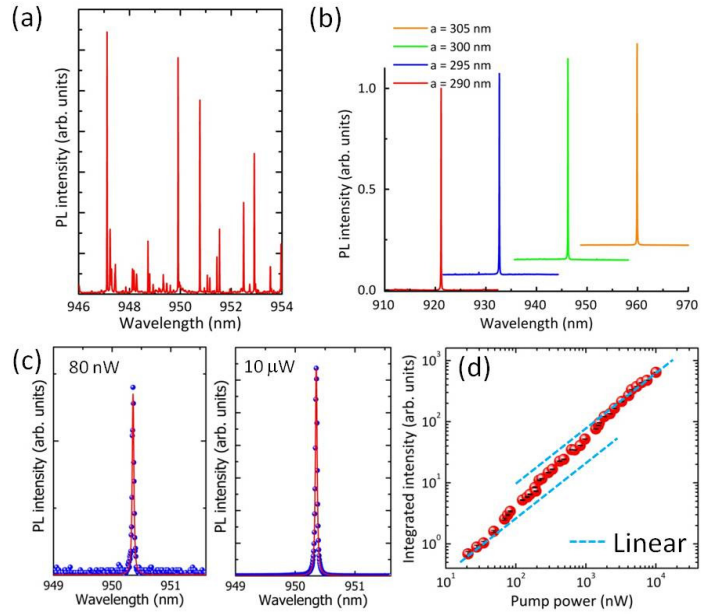

FIG. 2. (a) PL spectrum of QDs measured at 5 K in the unpatterned region. (b) PL spectra of the cavity modes with different lattice constants varying from 290 to 305 nm. The cavities show lasing with QD gain through the non-resonant coupling process. (c) PL spectra for an H0-type PhC nanocavity measured at 80 nW (left, below the threshold) and 10 μW (right, above the threshold). (d) L–L plot of the cavity mode. The laser threshold is ~300 nW with wetting layer excitation (880 nm).
5

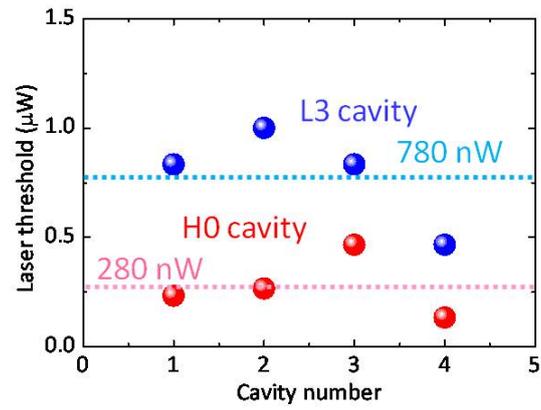

FIG. 3. Laser thresholds of four H0-type (red balls) and four L3-type (blue balls) nanocavities. The pink (280 nW) and light blue (780 nW) broken lines are the averaged values of the respective cavities. H0-type nanocavities have smaller thresholds than L3-type nanocavities.